\documentclass[a4paper,fleqn,usenatbib]{mnras}

\usepackage{amsmath,amssymb}

\usepackage[T1]{fontenc}
\usepackage{ae,aecompl}
\usepackage{graphicx}	
\usepackage{amsmath}	
\usepackage{amssymb}	
\usepackage{pifont}
\usepackage{txfonts}
\usepackage{url}
\usepackage{subfigure}
\usepackage{longtable}
\usepackage{lscape}
\usepackage{natbib}

\usepackage{graphicx}	
\usepackage{amsmath}	
\usepackage{amssymb}	
\newcommand{\mjyb}{{\rm mJy~beam}^{-1}}
\newcommand{\jyb}{{\rm Jy~beam}^{-1}}
\newcommand{\Msun}{\rm M_{\sun}}

\title[Gravitational collapse in turbulent media]{Mass-size scaling $M \sim
r^{1.67}$ of massive star-forming clumps -- evidences of turbulence-regulated gravitational collapse}

\author[C. P. Zhang et al.]{Chuan-Peng Zhang$^{1}$,
Guang-Xing Li$^{2}$\thanks{E-mail: gxli@usm.uni-muenchen.de}\thanks{Both authors contributed equally to this work.}\\
$^{1}$National Astronomical Observatories, CAS, 20A Datun Road, Chaoyang District, 100012 Beijing, China\\
$^{2}$University Observatory Munich, Scheinerstrasse 1, D-81679 Munich, Germany\\
}

\date{Accepted XXX. Received YYY; in original form ZZZ}

\pubyear{2016}

\begin{document}
\label{firstpage}
\pagerange{\pageref{firstpage}--\pageref{lastpage}}
\maketitle

\begin{abstract}
We study the fragmentation of eight massive clumps using data from ATLASGAL
870 $\mu$m, SCUBA 850 and 450 $\mu$m, PdBI 1.3 and 3.5 mm, and probe the fragmentation from
1 pc to 0.01 pc scale.  We find that the masses and the sizes of our objects
follow $M \sim r^{1.68\pm0.05}$. The results are in agreements with
the predictions of Li (2017) where $M \sim r^{5/3}$.  Inside each object, the
densest structures seem to be centrally condensed, with $\rho(r)\sim
r^{-2}$.
Our observational results support a scenario where molecular gas in the Milky Way is supported by a turbulence characterized by a constant energy dissipation
rate, and gas fragments like clumps and cores are structures which are
massive enough to be dynamically detached from the ambient medium.


\end{abstract}

\begin{keywords}
turbulence -- gravitation -- ISM: kinematics and dynamics -- instabilities-- galaxies: star formation
\end{keywords}


\section{Introduction}
Star formation is a fundamental process in the Milky Way and other galaxies. 
 The star-forming regions exhibit structures over multiple scales, and
they are believed to be primarily shaped by the interplay between turbulence and
gravity
\citep{2004RvMP...76..125M,2005ApJ...630..250K,2007ApJ...654..304K}.
Many diagnostics have been
developed to quantify the properties of star-forming regions, which enable one to compare
models with observations. These include the surface density probability
density distribution (PDF)
\citep{2009A&A...508L..35K,2011ApJ...727L..20K,2013MNRAS.436.1245F,2001ApJ...557..727V}, the correlation function
\citep{2009ApJ...707L.153P,2009A&A...504..883B,1999ApJ...524..887R}, as well as
the gravitational energy spectrum
\citep{2016MNRAS.461.3027L,2017MNRAS.464.4096L} which quantifies the
multi-scale distribution of gravitational energy.

Comparing masses and sizes is a straightforward way to study the properties
of the dense gas fragments. Previously, various mass-size relations have
been proposed in the literature.
Perhaps the most well-recognised one is the $M \sim r^{1.33}$ scaling by
\citet{Kauffmann2010} as a threshold condition for regions to form massive
 protostars. In this paper, we compare the properties of
 sub-millimetre bright gas fragments in the mass-size plane. We are
 interested in  sub-parsec scale structures that belong to the dense parts
 of the molecular interstellar medium (ISM), and the majority of them should
 collapse monotonically to form stars and star clusters. 
 
Although the $M \sim r^{1.33}$ scaling  by \citet{Kauffmann2010} has
been widely used to interpret observations, it does not offer
a description to the properties of the observed
gas fragments. In fact, the observed dense gas fragments seem to follow
a scaling relation with a steeper slope
\citep{Urquhart2014,2016A&A...585A.117Z,2016A&A...588A.104G}. Recently, it has
been suggested by \citet{Pfalzner2016} that both the cluster-forming
clumps \citep{Urquhart2014,2015A&A...579A..91W} and the embedded star clusters
\citep{2003ARA&A..41...57L}  obey $M \sim r^{1.67}$.
 \citet{2016arXiv161006577L} interprets this scaling as a result of
gravitational collapse regulated by ambient turbulence.
Although somewhat different scaling exponents have been reported in the
literature\footnote{E.g. a
 fit to a sample containing sources at different evolution stages by
 \citet{2015A&A...579A..91W} yields a steeper slope, where $M \sim r^{1.76\pm
 0.01}$, this can be compared with the \citet{Urquhart2014} result $M \sim
 r^{1.67\pm 0.036}$ where they selected only sources hosting compact H
 {\scriptsize {II}} regions.  The discrepancy thus comes
 from sample selection.  This can be seen from Fig. 24 of 
 \citet{2015A&A...579A..91W} where they plotted the mass-size relation of
 two sub-samples of sources at different evolutional stages.  The
 difference is mainly contributed from a few data points at 0.01 pc scale
 where one might suspect some selection effects. Since the sub-samples might
 have their own selection biases, the current ATLASGAL sample probably can not distinguish between $M \sim r^{1.67}$ and $M \sim r^{1.76}$.
  Further observational efforts are needed.}, most of the clumps
  follow a relation that is steeper than $M \sim r^{1.33}$ and is close to  $M \sim r^{1.67}$.

Due to limited resolution, previous analyses of the mass-size relation
of the gas fragments focus on parsec-scale structures.
On the other hand, according to \citet{2016arXiv161006577L}, properties
of the dense gas fragments are determined by the interplay between turbulence
and gravity over multiple scales, so it is natural to expect the mass-size relation
$M \sim r^{1.67}$ to extend down to smaller scales. In this paper, we study the
properties of gas fragments in star-forming regions by combining new
observations with literature data. The observations were carried out with the
Plateau de Bure Interferometer (PDBI) observational results with a criterion for
quasi-isolated gravitational collapse derived in \citet{2016arXiv161006577L}.

\section{Observations and data reduction}

\subsection{ATLASGAL clumps}

The eight massive precluster clumps (G18.17,
G18.21, G23.97N, G23.98, G23.44, G23.97S, G25.38, and G25.71) were
selected from the SCUBA Massive Pre/Protocluster core Survey
\citep[SCAMPS;][]{Thompson2005}.
These massive clumps have been covered by the Atacama Pathfinder Experiment
(APEX) Telescope Large Area Survey of the Galaxy \citep[ATLASGAL\footnote{The
ATLASGAL project is a collaboration between the Max-Planck-Gesellschaft, the
European Southern Observatory (ESO) and the Universidad de
Chile.};][]{Schuller2009}. \footnote{The data can be downloaded in the ATLASGAL
data base server \url{http://atlasgal.mpifr-bonn.mpg.de}. The ATLASGAL
samples have been described in detail by \citet{Csengeri2014}.
}
The survey was carried out at 870 $\mu$m with a beam
size of 19.2$''$ with a rms in the range 40 to 60 $\mjyb$, and had the goal of
studying cold clumps associated with high-mass star-forming regions.

\subsection{SCUBA clumps}

The eight clumps were observed by the Submillimetre
Common-User Bolometer Array (SCUBA) survey at 850 and 450
$\mu$m with James Clerk Maxwell Telescope
\citep[JCMT;][]{2013MNRAS.430.2513H,2013MNRAS.430.2545C} \footnote{The data can be downloaded in the JCMT science
archive \url{http://www.cadc-ccda.hia-iha.nrc-cnrc.gc.ca/en/jcmt/}.}. The
resolutions at 850 $\mu$m and 450 $\mu$m are $14''$ and $8''$ with rms of 0.20
and 1.6 $\jyb$, respectively.

\subsection{PdBI cores, and condensations}

The eight sources were observed with the IRAM\footnote{IRAM is supported by
INSU/CNRS (France), MPG (Germany) and IGN (Spain).} Plateau de Bure
Interferometer (PdBI) at 3.5 and 1.3 mm simultaneously with CD 
(for all eight sources) and BCD
(only for four dense sources G23.44, G23.97S, G25.38, and G25.71)
configurations during 2004 and 2006. The 3.5 mm receivers were tuned to 86.086
GHz in single sideband (SSB) mode. The 1.3 mm receivers were tuned to 219.560
GHz in double side-band (DSB) mode. Both receivers used two 320 MHz wide
backends for continuum observation. For calibrations, the quasar B1741-038 was
used as phase calibrator, and the quasar 3C273 and the evolved star MWC 349 were
used as flux calibrators.

The CLIC and MAPPING modules in IRAM software package
GILDAS\footnote{http://www.iram.fr/IRAMFR/GILDAS/} were used for
the data calibration and cleaning. The primary (synthesis) beam
is about 58.5$''$ ($4.2'' \times 2.8''$) at 86.086 GHz (or 3.5 mm)
of CD tracks with a sensitivity of about 0.30 $\mjyb$, and about
23.0$''$ ($0.8'' \times 0.6''$) at 219.560 GHz (or 1.3 mm) of BCD
tracks with a sensitivity of about 0.65 $\mjyb$, respectively.
Details concerning the PdBI data will be presented
in forthcoming paper (Zhang et al. 2017 in prep.).

\section{Results}

\subsection{Source extraction}
\label{sect_definition}

We adopt typical terminologies where clumps, dense cores, and condensations
as structures of physical FWHM sizes of $\sim$1, $\sim$0.1, and $\sim$0.01
pc, respectively \citep{Williams2000}. The \textit{Gaussclumps}
\citep{Stutzki1990,Kramer1998,Csengeri2014} in
the GILDAS software package was used to
structures at different scales and wavelengths
from our samples. The \textit{Gaussclumps} fits 2-dimensional gaussians
locally to the maximums of the input data cube. It then subtracts this fragment from the
cube, creates a residual map, and continues with the maximum of this
residual map.
The procedure is repeated until a stop criterion is met,
for instance when the maximum of the residual maps drops below a certain
level. In this paper, fragments with peak intensity above 5$\sigma$ are
considered as signals. 

\subsection{Clump mass estimation}
\label{sect_mass}

It is assumed that the dust emission is optically thin and the-gas
to-dust ratio is 100 \citep{1991ARA&A..29..581Y}. The dust temperature is
estimated using NH$_3$ (1, 1) and (2, 2) rotational lines observed by the Very
Large Array (Zhang et al. 2017 in prep.).
The fragment masses are calculated using dust opacities 0.002
cm$^2$ g$^{-1}$ at 3.5 mm, 0.009 cm$^2$ g$^{-1}$ at 1.3 mm,
and 0.0185 cm$^2$ g$^{-1}$ at 870 $\mu$m
\citep{Ossenkopf1994,Pillai2011}.
The total masses $M$ of the sources can be calculated
 using \citep{Kauffmann2008}
	\begin{eqnarray}
	\begin{aligned}
	\label{equa_mass}
	\left(\frac{M}{\Msun}\right)   =  & 0.12 \left({\rm  e}^{14.39\left(\frac{\lambda}{\rm mm}\right)^{-1}\left(\frac{T_{\rm dust}}{\rm K}\right)^{-1}}-1\right) \times &  \\
        & \left(\frac{\kappa_{\nu}}{\rm cm^{2}g^{-1}}\right)^{-1} \left(\frac{S_{\nu}}{\rm Jy}\right) \left(\frac{D}{\rm kpc}\right)^{2}
	\left(\frac{\lambda}{\rm mm}\right)^{3}, &
	\end{aligned}
	\end{eqnarray}
where $\lambda$ is the observational wavelength, $T_{\rm dust}$
is the dust temperature, $\kappa_{\nu}$ is the dust opacity,
$S_{\nu}$ is the integrated flux, and $D$ is the distance to
the Sun. The integrated flux at 3.5 mm may be contributed by both dust and
free-free emissions which were distinguished
using spectral energy distribution (SED) fitting to multi-wavelength continuum
data \citep[e.g.][]{Zhang2014}.
The detailed procedure and results will be presented in Zhang
et al. (2017 in prep.). The mass uncertainties are mainly contributed from the
uncertainties of the dust temperature estimates.

\subsection{The results of ATLASGAL, SCUBA, and PdBI observations}

\begin{figure*}
\centering
\includegraphics[width=0.70\textwidth, angle=0]{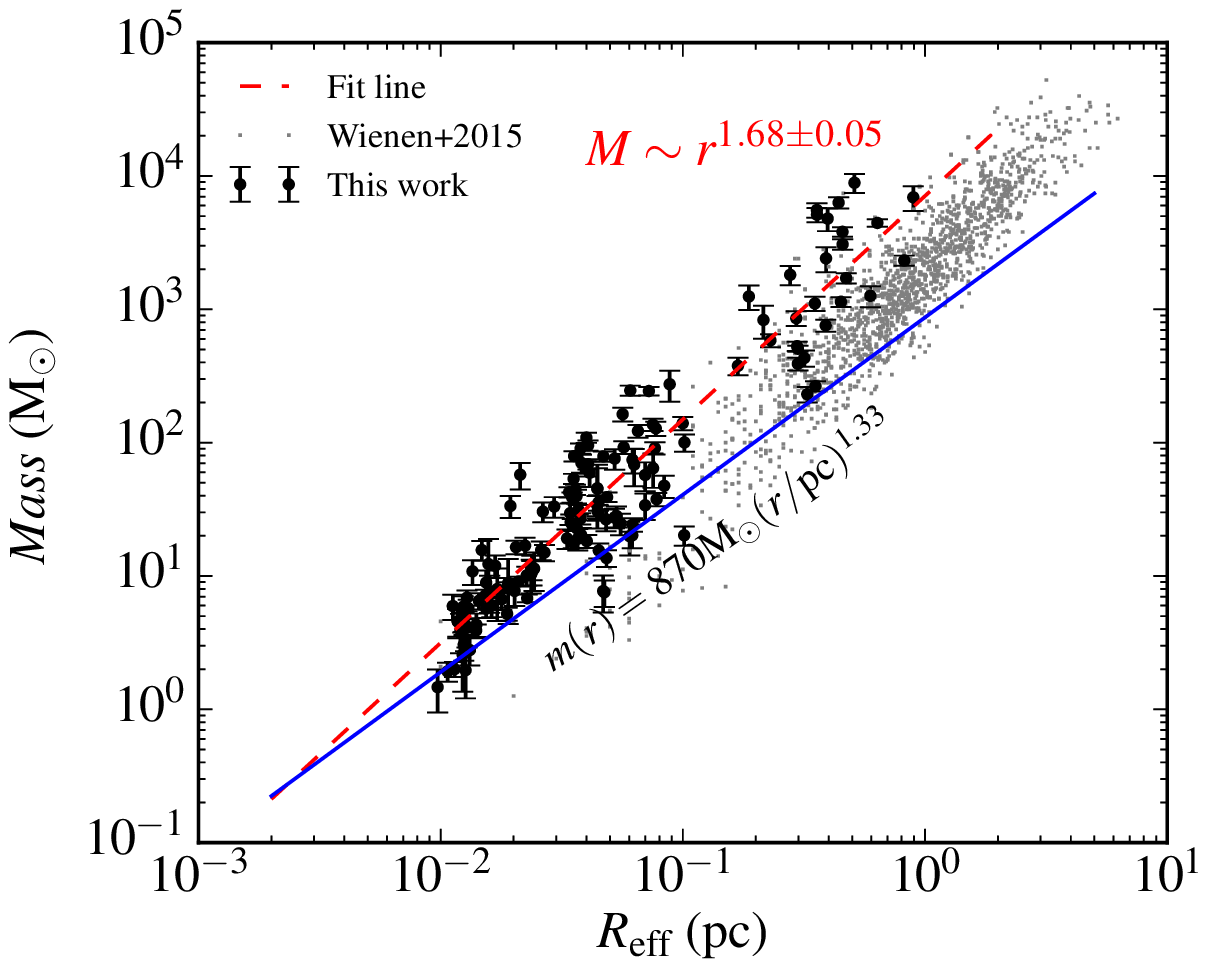}
\caption{Mass-size plane for the eight sources (G18.17,
G18.21, G23.97N, G23.98, G23.44, G23.97S, G25.38, and G25.71) at
different wavelengths, together with the clumps from \citet{2015A&A...579A..91W}.
The masses are derived from the integrated flux within a measured
Gaussian $FWHM$ using the \textit{Gaussclumps}, and the effective radii are
defined as $R_{\rm eff} = FWHM/(2\sqrt{\rm ln2})$. The straight blue line
shows a threshold to form high-mass protostars
\citep{Kauffmann2010}. The dashed red line shows the result of
a power-law fit to the whole sample, where we find
${\rm log}(Mass(R_{\rm eff})/{\Msun}) = 3.85 + 1.68 {\times}
{\rm log} (R_{\rm eff}/ {\rm pc})$ or $M \sim r^{1.68\pm0.05}$
for short. It is comparable with the analytical result
of \citet{2016arXiv161006577L} where $M \sim r ^{5/3}\sim  r^{1.67}$.}
\label{Fig:mass-size-whole}
\end{figure*}

In Figure \ref{Fig:mass-size-whole} we plot the structures in the mass-size
plane from 1 pc to 0.01 pc scale,  by combining 870 $\mu$m, 850
$\mu$m, 450 $\mu$m, 3.5 mm, and 1.3 mm observational data (including
data from G18.17, G18.21, G23.97N, G23.98, G23.44, G23.97S, G25.38, and
G25.71). For comparison, we also plot the ATLASGAL clumps in
\citet{2015A&A...579A..91W}. Our our sample sources are not included in their
sample. The mass-size threshold $m(r) > 870 {\rm \Msun} (r/{\rm pc})^{1.33}$ for
massive star formation \citep{Kauffmann2010} is also plotted for reference. A power-law fit to our data yields ${\log}(\rm Mass(R_{\rm eff})/{\Msun}) = 3.86 + 1.68
{\times} {\rm log}(R_{\rm eff}/ {\rm pc})$ or $(M/7000{\rm M_{\odot}}) \approx
(r/{\rm pc})^{1.68\pm0.05}$ for short. The slope we obtained is similar
to what was found in \citet{Urquhart2014} and \citet{2015A&A...579A..91W}.
Our observations are selected to target at the most prominent fragments,
 which is probably the reason that the normalization mass-size scaling is
higher than those presented in \citet{Urquhart2014} and
\citet{2015A&A...579A..91W}.

\begin{figure*}
\centering
\includegraphics[width=0.49\textwidth, angle=0]{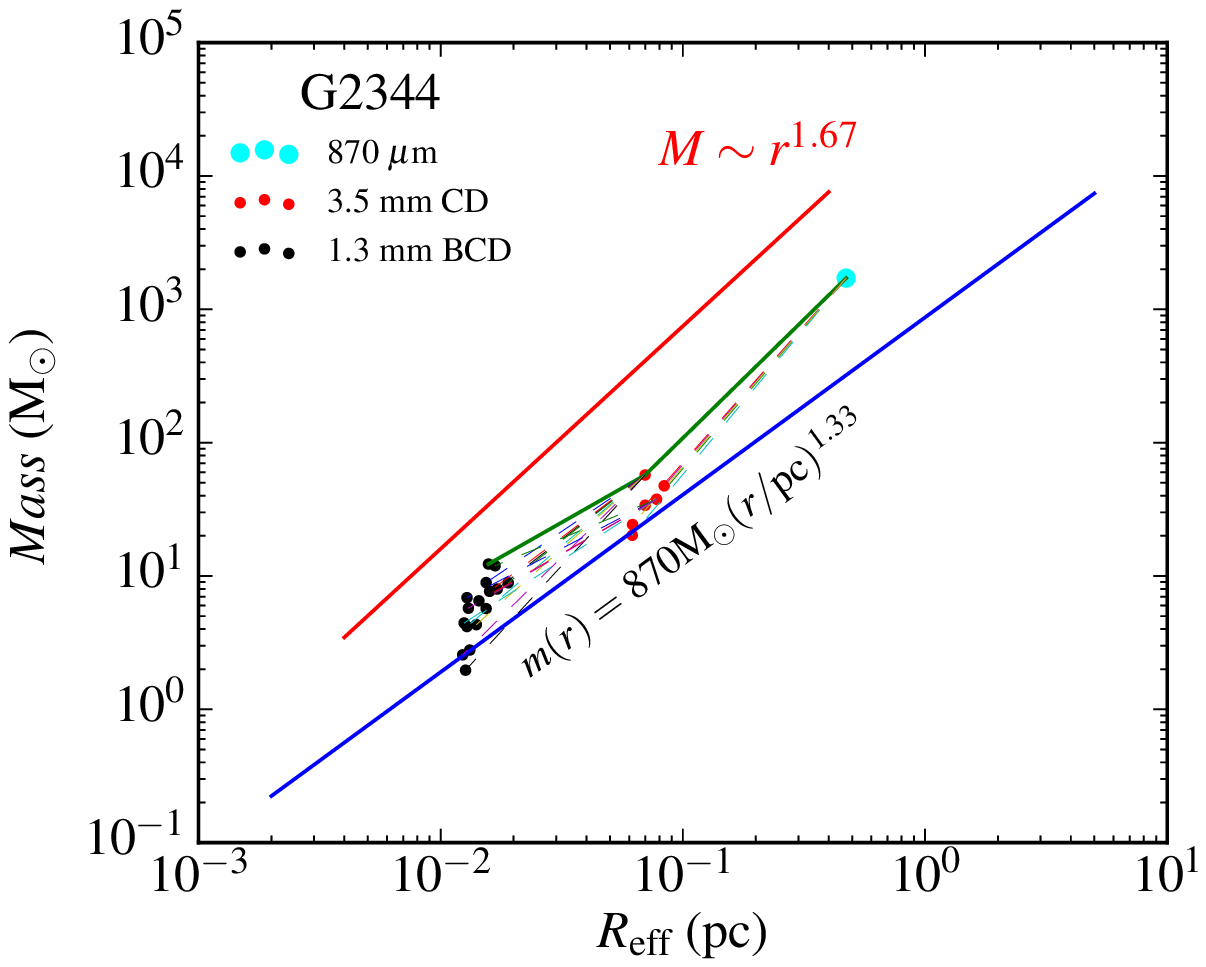}
\includegraphics[width=0.49\textwidth, angle=0]{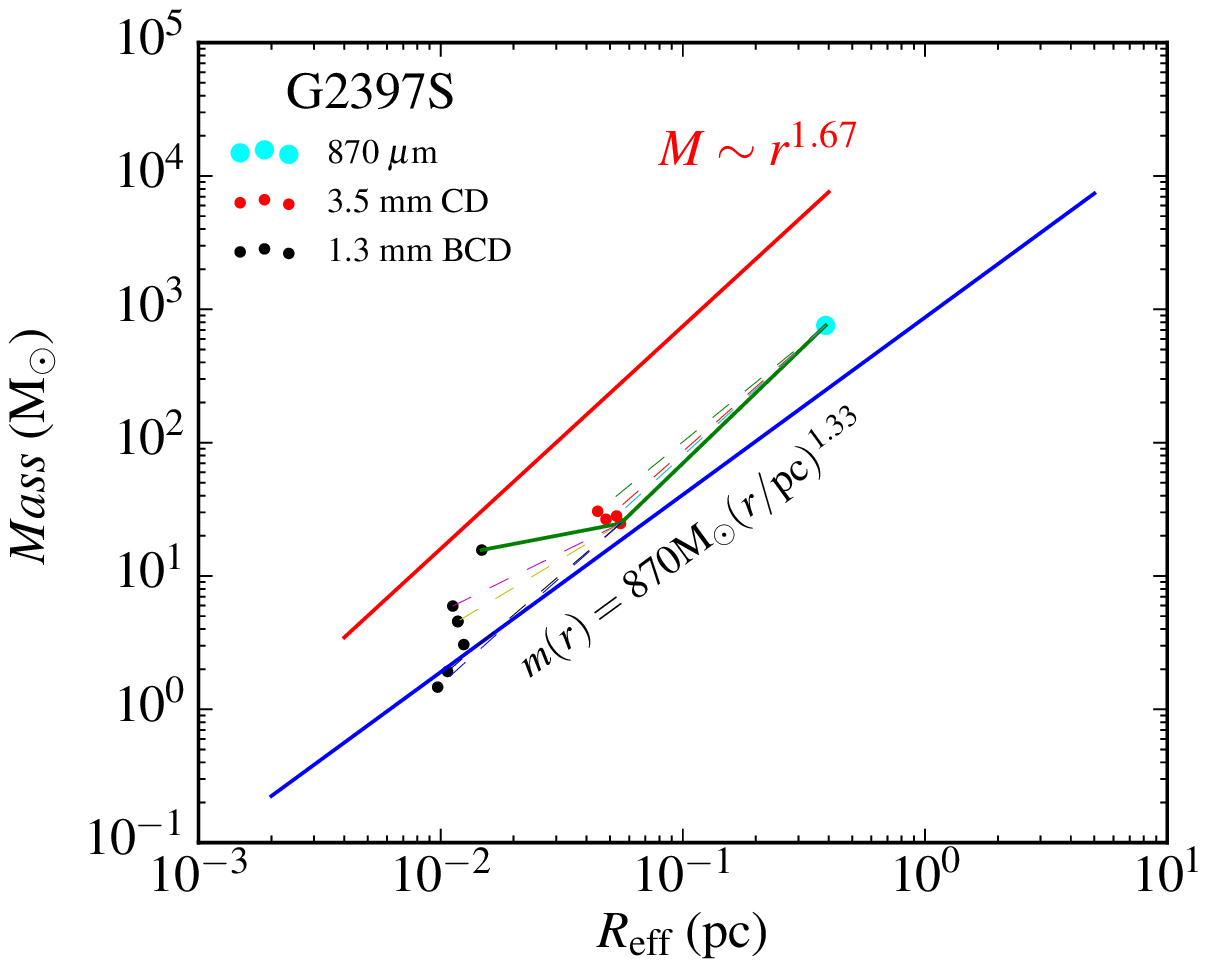}
\includegraphics[width=0.49\textwidth, angle=0]{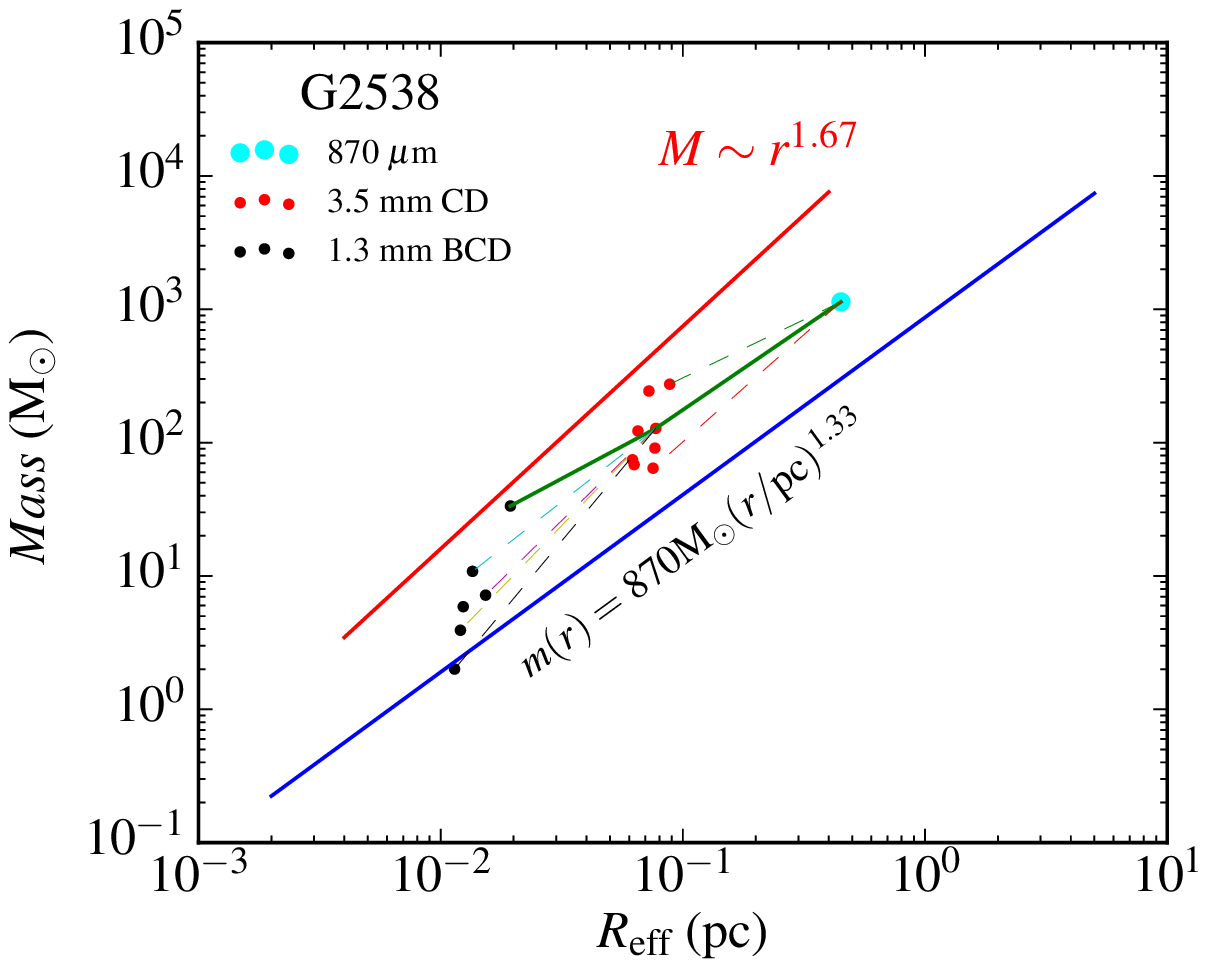}
\includegraphics[width=0.49\textwidth, angle=0]{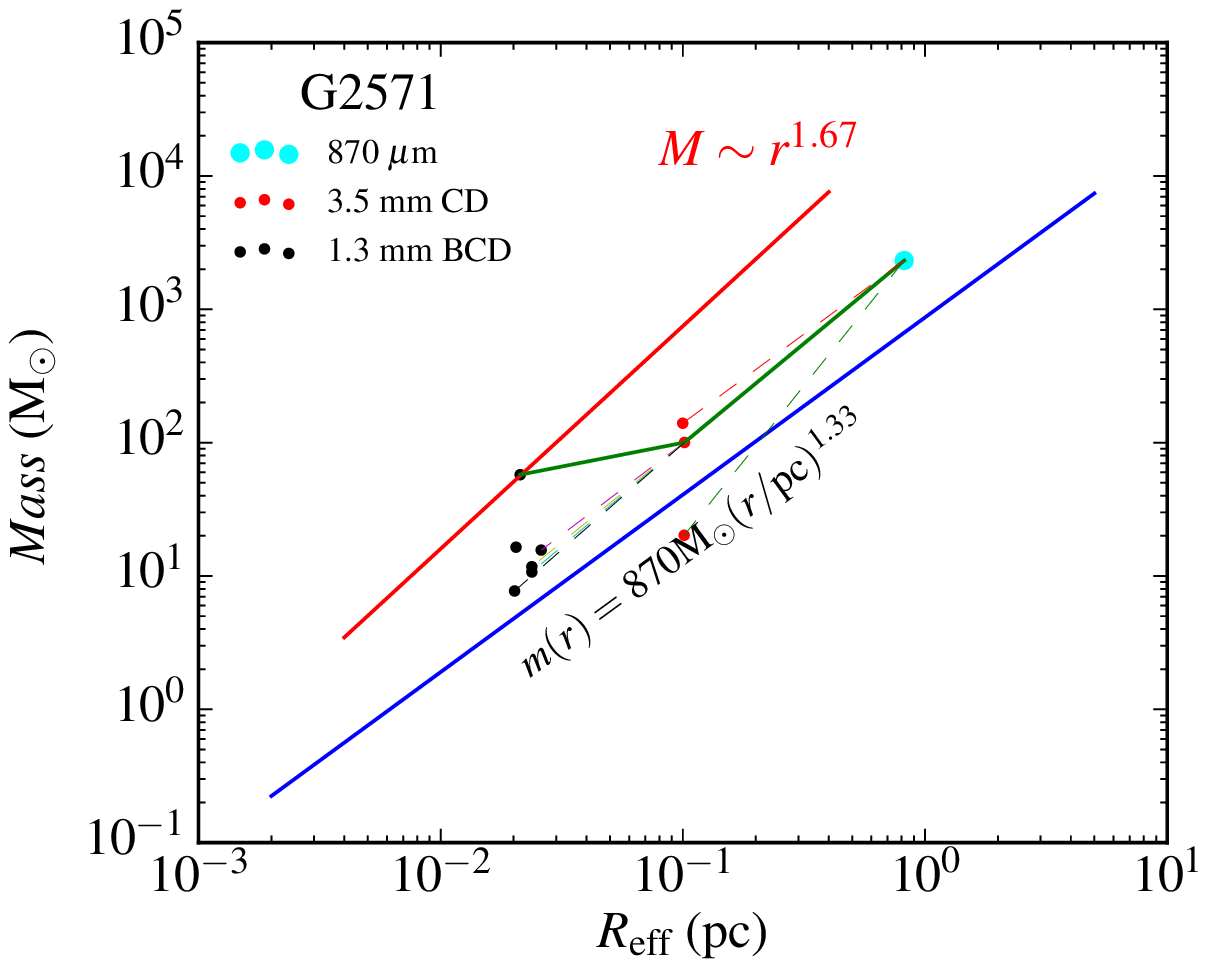}
\caption{Fragmentation diagram of individual sources in the mass-size
plane using data obtained at 870 $\mu$m, 3.5 mm CD track, and 1.3 mm BCD track.
The masses are derived from the integrated flux within a measured Gaussian
profile derived using the \textit{Gaussclumps}, and the effective radius is
$R_{\rm eff} = FWHM/(2\sqrt{\rm ln2})$.
The straight blue lines show a threshold to form high-mass protostars
\citep{Kauffmann2010}. The thick green lines represent the set of structures within
which the most prominent cores reside. The straight red line shows the scaling
$M \sim r^{1.67}$ derived in \citet{2016arXiv161006577L} as the threshold
condition for gravitational collapse in a turbulent medium.}
\label{Fig:mass-size-individual}
\end{figure*}

In Figure \ref{Fig:mass-size-individual} we plot the properties of the
individual regions viewed in the mass-size plane. Different color-filled
circles stand for the clumps and cores, respectively. We use
dashed lines to connect the condensations with their host cores or clumps. Finally,
we draw the thick green lines to connect the spatially overlapping
objects that are most prominent at a given scale \footnote{In many cases, we
can identify significant free-free emission components from the
multi-wavelength observations of these clumps. These sources are more evolved
objects, and tend to host more massive dense cores.}.
These tree-diagrams enable us to represent the nested
structure of a given clump. In some sense this is similar to a Dendrogram
representation of the data \citep{2008ApJ...679.1338R}.

We should note that the different observations have different spacial coverages:
due to the limitation of the size of the primary beams of interferometer array,
the 1.3 mm high-resolution observations cover only the central (23.0$''$) part of the
region observed at 870 $\mu$m.  We are only able to probe
the fragmentation down to about 0.01
pc for four sources where the 1.3 mm high-resolution observations are
available.

\begin{figure*}
\centering
\includegraphics[width=0.50\textwidth, angle=0]{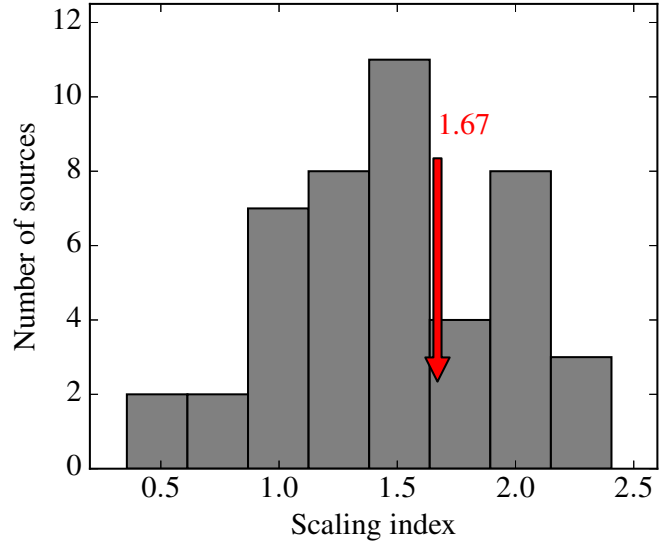}
\caption{Histogram for the scaling index. The $x$-axis is the scaling index $k_{\rm m}$
of the mass-size relation ($M \sim r^{k_{\rm m}}$), the $y$-axis is the number
of sources in each bin. A bin size of 0.26 dex is used. The distribution has a
mean of 1.47, median of 1.48, and a standard deviation of 0.48. The red arrow
points to the threshold condition $M \sim r^{1.67}$ for gravitational collapse
derived in \citet{2016arXiv161006577L}.}
\label{Fig:histogram}
\end{figure*}

In Figure \ref{Fig:histogram} we plot the distribution of the scaling
exponents of all the pairs of nested structures. The red arrow points to the
threshold condition $M \sim r^{1.67}$ \citep{2016arXiv161006577L} for the model
of turbulence-regulated gravitational collapse.

\section{Discussion: Fragmentation in the mass-size plane}

\subsection{Gravitational collapse in a turbulent medium}
Gravitational collapse in turbulent media has been intensively studied
before. Earlier treatments such as
\citet{1951RSPSA.210...26C} and \citet{1952Natur.170.1030P} assume constant velocity
dispersions for the turbulent gas. In today's view, their picture is
probably over-simplified, as the multi-scaled structure of the turbulence flow is not
well characterized. Subsequent works \citep{1987A&A...172..293B,1995A&A...303..204V}
describe the instability in a framework where they assume Kolmogorov-like power
spectrum of turbulent energy and fractal-like density-size scalings. In
particular, \citet{1995A&A...303..204V} identified different regimes within
which the interplay between turbulence and gravity produces different modes of
collapse. The study revealed a diverse range of possibilities where
gravitational collapse would occur. These results are interesting theoretically.
However, a remaining difficulty is to study the non-linear outcome of the
instability and to relate the structures they produced in the Fourier
space with the dense gas fragments seen in observations.
This task is highly non-trivial \citep{1990ApJS...72..133H}.

Following previous studies,
\citep{2010ApJ...716..433K,2011MNRAS.411...65B,2012A&A...545A.147H,2016A&A...591A..30L},
\citet{2016arXiv161006577L} considered the evolution of a dense object embedded in a diffuse medium, and study the interaction between the dense object and
the ambient medium. They assume a picture where  inside the object, gravity is driving
contractions, and outside the object, turbulence is providing support.   They carried out the analysis in the
physical space, and studied the outcome of the interaction between the object
and the ambient medium. Their main conclusion is that when the turbulence energy dissipation rate of the ambient medium is close to a constant,
the objects that undergo gravitational collapse should follow the scaling relation
$M \sim r^{1.67}$.

Note that the condition is independent on how objects are
formed. It is certainly possible that the structures we see originate from a
hierarchical, bottom-up assembly of smaller structures, as has been proposed by
\citet{2011MNRAS.411...65B,2011MNRAS.416.1436B} and \citet{2016arXiv161100088V}.
As long as the ambient have an almost-uniform level of turbulence, the scaling $M \sim
r^{1.67}$ should hold.

\subsection{$M \sim r^{1.67}$ as the threshold condition for gravitational collapse}
We use the analytical prediction developed in
\citet{2016arXiv161006577L} to interpret our results. In their picture, the
molecular ISM consists of two dynamical phases -- a diffuse phase where
turbulence dominates and a dense phase where gravity dominates. One major
prediction of \citet{2016arXiv161006577L} is that if the ambient turbulence is
characterized by a constant energy dissipation rate $\epsilon$, gravitationally
bound structures should obey $M \approx \epsilon^{2/3} \eta^{−2/3}G^{−1}r^{5/3}$
where $G$ is the gravitational constant and
$r$ is the size. $\eta$ is a parameter for turbulent dissipation, and is close
to unity. Thus the mass-size scaling is determined by the energy dissipation
rate of the ambient medium $\epsilon$. When $\epsilon$ is a constant, $M \sim
r^{5/3}$. This scaling has been supported by data from \citet{Pfalzner2016}.

Gas condensations of smaller sizes were not studied in \citet{Pfalzner2016},
because they are limited by the resolution of the single-dish continuum
observations. However, the theoretical limit $M \sim r^{5/3}$ was derived by
considering the interaction between gravity and an ambient turbulence, and one
expects it to hold as long as turbulence can cascade effectively. In the
molecular ISM, the limit of this cascade, namely the Kolmogorov microscale is
very small ($10^{-6}$ pc as estimated in \citet{1983ApJ...272L..45F}). Thus
we expect this relation to be valid for structures on smaller scales.

With our new data we are probing structures down to $\sim 0.01 \,\rm pc$ scale.
Fig. \ref{Fig:mass-size-whole} plots the mass-size plane from clump scale
to $\sim 0.01\;\rm pc$ condensation scale, and a fit to the observational data suggests an
universal mass-size relation  ${\rm log} ({\rm Mass}/{\Msun}) = 3.85 + 1.68 {\times}
{\rm log} (R_{\rm eff}/ {\rm pc})$ (or $M \sim r^{1.68\pm0.05}$ for short)
that is valid from a few parsec to $\sim 0.01$ pc. Thus the scaling
relation $M \sim r^{5/3} \sim r^{1.67}$ offers a good explanation to our
multi-scale observational data. According to
\citet{2016arXiv161006577L},  this is a direct consequence of the fact that the
diffuse phase of the Milky Way ISM is dominated by a turbulence with a constant
energy dissipation rate, and the observed structures are the dense parts that
are dynamically detached from the ambient medium and are probably collapsing.

Other attempts have been made to explain the observed mass-size relation.
It is straightforward to combine the Larson's relation with the
virial equilibrium to derive similar mass-size scalings. This has been
previously attempted
\citep{2010ApJ...716..433K,2011MNRAS.411...65B,2012A&A...545A.147H,2016A&A...591A..30L}
\footnote{Although all these authors would agree that a combination of
virial equilibrium with the Larson's relation would produce a mass-size relation,
the underlying pictures are different. \citet{2010ApJ...716..433K} used
the Larson's relation to make estimates, and did not specify a physical picture.
\citet{2011MNRAS.411...65B} believe both the mass-size relation and the
Larson's relation would arise from gravitational collapse.
\citet{2016A&A...591A..30L} and \citet{2016arXiv161006577L} believe that
turbulent is regulating the gravitating collapse, and
\citet{2016arXiv161006577L} invoked the constant energy dissipation
rate argument to analytically derive the $M \sim  r^{1.67}$ relation. }.
The derived scalings are somewhat similar to the one we observed. However, the
interpretation of \citet{2016arXiv161006577L} seems to be preferred over the
previous ones for two reasons: first, the Larson relation has its own
uncertainties; in the relation, the velocity dispersion $\sigma_{\rm v}$
and scale $l$ are related by $\sigma_{\rm v} \sim l^{\alpha}$ where the value of $\alpha$ is found to range from 1/3 to 1/2
\citep{1981MNRAS.194..809L,2009ApJ...699.1092H, 2011ApJ...740..120R}. If one
simply derives the mass-size relation based on the Larson's relation,
assuming that the dense structures are gravitationally bound, using
$M \sim \sigma_{\rm v}^2 r / G$ one would predict a variety of mass-size
scalings range from $M \sim r^{1.67}$ to $M \sim r^{2}$, and  on the other
hand, the observed scaling  $M \sim r^{1.68\pm0.05}$ 
stays very close to the $M \sim
r^{1.67}$ prediction made by \citet{2016arXiv161006577L}. Second, it has
been found that the clumps themselves do not obey the Larson's relation
\citep[e.g. appendix A1 of][]{2015A&A...579A..91W}. It is the ambient medium
 that satisfies the relation. Therefore, one should not directly combine
the Larson's relation with the gravitational bound argument to derive the mass-size relation.
Realizing this, \citet{2016arXiv161006577L} did not invoke the assumption
that the dense structures  are themselves following the Larson's relation.
Instead, they assume a constant energy dissipation rate in the ambient
turbulence, and study the interaction between the object and the ambient
turbulent medium. It is this interplay that
determines the mass-size relation.
In their picture, the ambient medium is dominated by turbulence, and follows the
Larson's relation. The objects are dynamically detached from the ambient medium,
and they do not have to obey the relation.

\subsection{Slopes in the mass-size plane}

When an object is dynamically detached from the ambient medium, its evolution
should be determined by the dynamics of the gravity-driven turbulence inside the
object. Dedicated numerical simulations as well as analytical models have been
developed to describe the collapse. When a region undergoes gravitational
collapse, the density profile approaches $\rho \sim r^{-2}$
\citep{2007ApJ...665..416K,2010A&A...512A..81F,2015ApJ...804...44M}. However,
one should note that this density profile can also be produced by gravitational
free-fall \citep{2014ApJ...781...91G} and magnetized models
\citep{2007ApJ...671..497A}.

In observations, many of the star-forming regions seem to have $\rho\sim
r^{-2}$. This should correspond to $M \sim r$. In Fig.
\ref{Fig:mass-size-individual} we plot the properties of different gas fragments
in the mass-size plane, with connecting lines in the
plot to denote structures that are  spatially overlapping on the map. In
Fig.
\ref{Fig:histogram} we present the distribution the slopes of these connecting
lines in the mass-size plane. Since the high-resolution observations only cover
the most central (23$''$) parts of the regions, we are essentially probing
only the densest fragments. The massive connected fragments
are much more centrally condensed than
$M \sim r^{5/3}$ (31 of the connected structures out of 45 have slopes that are
shallower than $M \sim r^{1.67}$) and  seems to approach $\rho(r)\sim
r^{-2}$, indicating that these structures are centrally-condensed enough to be dynamically detached from the ambient turbulence.

Using a technique called inverse dynamical population synthesis,
\citet{2012A&A...543A...8M} derived a mass-size relation $r/{\rm pc} \sim
(M/M_{\odot})^{0.13 \pm 0.04}$ for ``cluster-forming cloud clumps''.
Essentially, what they are constraining are the birthplaces of binary stars in
star clusters. Their results indicate that the binaries in a star clusters are
born within a very small radius (typically $\sim 0.1\;\rm pc$), and the scale is
only weakly dependent on the mass. This perhaps suggests that the fragments we
observed are still undergoing significant infall, and due to this transport, the
stars are likely to be born at the very centres of the clumps. Although
we can not explain this scaling analytically, it is at least consistent with our
picture where gas inside the clumps is dynamically
detached from the ambient turbulence and is collapsing.

\section{Conclusions}

In this work, we study the fragmentation of eight massive clumps using data from
ATLASGAL 870 $\mu$m, PdBI 1.3 and 3.5 mm, and probed the fragmentation from 1 pc
to 0.01 pc scale. Combined with previous measurements and analytical arguments,
we propose a mass-size relation $M \sim r^{1.67}$ that holds from around 0.01 pc
to 1 pc scale. The mass-size relation can be understood if the structures
undergo quasi-isolated gravitational collapse in a turbulent medium, as
predicted by \citet{2016arXiv161006577L}. The structures at the centres of the
clumps are more centrally-condensed and seem to approach $M \sim r$ ({which
is equivalent to $\rho(r)\sim r^{-2}$}).

Our observational results with $M \sim r^{1.68\pm0.05}$ support a scenario where
molecular gas in the Milky Way is supported by a turbulence with  an almost
constant energy dissipation rate, and gas fragments like clumps and cores are
structures which are dense enough to be dynamically detached from the ambient
medium.

\section*{Acknowledgements}
\addcontentsline{toc}{section}{ACKNOWLEDGEMENTS}

This work is partly supported by the National Key Basic Research
Program of China (973 Program) 2015CB857100 and National Natural Science
Foundation of China 11363004 and 11403042. Chuan-Peng Zhang is
supported by the Young Researcher Grant of National
Astronomical Observatories, Chinese Academy of Sciences. Guang-Xing Li is
supported by the Deutsche Forschungsgemeinschaft (DFG) priority program 1573
ISM-SPP. The James Clerk Maxwell Telescope has historically been
operated by the Joint Astronomy Centre on behalf of the Science and
Technology Facilities Council of the United Kingdom, the National
Research Council of Canada and the Netherlands Organisation for Scientific Research.
Additional funds for the construction of SCUBA-2 were provided by the
Canada Foundation for Innovation. Guang-Xing Li thanks Enrique Vazquez-Semadeni
for a thorough discussion on gravitational instability, and thanks Pavel Kroupa
for a discussion on star cluster formation.
Finally, the referee must acknowledged for the careful reports.








\bibliography{references}

\onecolumn

\label{lastpage}
\end{document}